\documentclass[aps,groupedaddress,10pt]{revtex4}
\usepackage{graphics}
\usepackage{graphicx}
\usepackage{dcolumn}
\usepackage{bm}

\newcommand{\de}{\partial}
\newcommand{\sech}[1]{\textrm{sech}\left(  #1\right)}
\newcommand{\eq}[2]{\begin{equation} \label{#1} #2 \end{equation}}

\newcommand{\etal}{{\em et al.}}

\begin{document}

\title{Theory of Raman multipeak states in solid-core photonic crystal fibers}

\author{Truong X. Tran$^{1}$, Alexander Podlipensky$^{1}$, Philip St.J. Russell$^{1}$ \& Fabio Biancalana$^{1}$}
\affiliation{$^{1}$Max Planck Institute for the Science of Light, G\"{u}nther-Scharowsky str. 1, Bau 26, 91058 Erlangen, Germany}

\begin{abstract}
Pulse splitting is a crucial and common process in nonlinear fiber optics. When an intense laser pulse is launched into a highly nonlinear fiber, a stream of fundamental solitons is generated, their temporal separations increasing during propagation. This is due to the onset of a variety of perturbations, including higher-order dispersion and the Raman effect. Recently, it has been experimentally observed that the well-known law determining the amplitudes and the temporal widths of each soliton, however, breaks down due to the unexpected formation of metastable 2-peak localised states with constant temporal separation between the two maxima. In the vicinity of certain 'magic' input powers the formation of 2-peak states is quite common in many types of highly nonlinear photonic crystal fibers. In this study, we provide a full theoretical understanding of the above recent observations. Based on a 'gravity-like' potential approach we derive simple equations for the 'magic' peak power ratio and the temporal separation between pulses forming these 2-peak states. We develop a model to calculate the magic input power of the input pulse around which the phenomenon can be observed. We also predict the existence of exotic multipeak states that strongly violate the perturbative pulse splitting law, and we study their stability and excitation conditions.
\end{abstract}

\maketitle

Pulse splitting is a well-known process occurring in the very initial moments of pulse propagation in nonlinear optical fibers \cite{splitting1,splitting2}. According to the most accredited theory of pulse splitting \cite{splitting1,splitting2}, in the femtosecond regime, higher-order solitons are affected by stimulated Raman scattering (SRS) and higher-order dispersion terms, becoming unstable and eventually breaking up into several fundamental solitons. Explicit expressions for the peak power $P_{j}$ and temporal width $T_{j}$ of the $j$-th fundamental soliton created in the splitting process are given by  ($1\leq j \leq N$):
\begin{eqnarray}
P_{j}=P_{0}(2N-2j+1)^2/N^2, \label{power}\\
T_{j}=T_{0}/(2N-2j+1), \label{width}
\end{eqnarray}
where $P_{0}$ and $T_{0}$ are the peak power and temporal width of the initial hyperbolic secant pulse and $N$ is the so-called soliton order, see Refs. \cite{splitting1,splitting2}. After splitting, intrapulse SRS causes fundamental solitons to shift continuously to lower frequencies via the Raman soliton self-frequency shift (RSFS) \cite{mitschke,gordon}. The rate of RSFS is proportional to  $1/T_{j}^4$ \cite{mitschke,gordon}. Thus, pulse splitting together with RSFS eventually leads to the complete breakup of a higher-order soliton, which ejects a stream of fundamental solitons one after another. According to Eqs. (\ref{power}-\ref{width}), solitons that are ejected earlier have higher amplitudes, shorter duration, and, as a result, demonstrate stronger RSFS. This leads to a sequence of fundamental solitons with different carrier frequencies, which constantly increase their temporal separations during propagation due to the RSFS-induced walk-off effect.

However, this well-known description ignores some processes that could affect the soliton dynamics, such as interaction between solitons, and the emission of dispersive wave radiation near a zero group-velocity dispersion (GVD) point. In a series of recent experiments (see Refs. \cite{pod1,pod2}), short and intense pulses have been launched in highly nonlinear photonic crystal fibers (PCFs, see also Ref. \cite{russell}). The unexpected formation of long-lived, 2-peak soliton states for specific 'magic' input pulse energies was observed \cite{pod1,pod2}. In these experiments, it was clear that this phenomenon is quite universal for a large variety of highly nonlinear optical fibers, especially in the deep anomalous GVD regime far from zero GVD points. Such localised states always show peaks of differing amplitude with a well-defined peak power ratio ($r\simeq 0.73$ in the mentioned experiments). Moreover, the temporal separation between the two peaks is also uniquely determined by $r$ and the Raman parameter $\tau_{R}$. Thus, under certain circumstances a pair of solitons with unequal amplitudes can form a localised state, which can propagate over long distances even in the presence of Raman effect.

Such soliton states were numerically discovered and studied in 1996 by Akhmediev \etal \cite{akhmediev}. Single peak and 2-peak states were found by solving the simplified nonlinear Schr\"odinger equation (NLSE), see Eqs. (\ref{NLSE}) - (\ref{Gagnon})  below. The analysis in Ref. \cite{akhmediev} is limited to two aspects: (i) finding the localized 2-peak solutions of Eq. (\ref{Gagnon}), and (ii) the question of numerical stability of such solutions.

In this paper, inspired by the recent experimental observation of the Raman multipeak states in Refs. \cite{pod1,pod2} and by the early numerical findings in Ref. \cite{akhmediev}, we investigate the issue of Raman multipeak states in detail. We propose qualitative and quantitative explanations for all the experimental observations, based on the powerful concept of gravity-like potential introduced in Refs. \cite{belanger,akhmediev,gorbach,gorbachnature}. The general point-of-view introduced by our approach allows one to make several important predictions to be tested in future experiments, namely the possibility of forming Raman soliton states with more than two peaks in PCFs, thus leading to a complete violation of the soliton splitting law of Eqs. (\ref{power}-\ref{width}), and to new ways to manipulate SCG in microstructured fibers by controlling exotic states of light in the fiber.

\section{Results}
\label{results}

\subsection{Multipeak solitons}
\label{multipeaksolitons}

A simplified form of NLSE can be written, in dimensionless units, as
\eq{NLSE}{i\de_{z}\psi+\frac{1}{2}\de_{t}^{2}\psi+|\psi|^{2}\psi-\tau_{R}\psi\de_{t}|\psi|^{2}=0,} where $\psi$ is the electric field envelope rescaled with the fundamental soliton power $P_{S}\equiv |\beta_{2}|/(\gamma t_{0}^{2})$, $\beta_{2}$ being the GVD coefficient at the reference frequency. Variables $z$ and $t$ are dimensionless space and time, respectively rescaled with the second order dispersion length $L_{\rm D2}\equiv t_{0}^{2}/|\beta_{2}|$ and with the input pulse duration $t_{0}$. The last term of Eq. (\ref{NLSE}), responsible for the Raman effect, produces a constant RSFS for solitons in silica fibers \cite{mitschke,gordon}, $\tau_{R}\equiv T_{R}/t_{0}$ being a small parameter, where $T_{R}$ is the Raman response time, approximately equal to $2$ fs in silica. Solitons subject to the Raman effect shift continuously towards the red part of the spectrum due to RSFS, leading to a constant acceleration of the soliton in the time domain \cite{mitschke,gordon}.
In the reference frame of an intense accelerating soliton, one can operate the Gagnon-B\'elanger transformation $\psi(z,t)=f(\xi)\exp\left[iz\left(q-b^{2}z^{2}/3+bt\right)\right]$, where $\xi\equiv t-bz^{2}/2$ and $b=32\tau_{R}q^{2}/15$. This leads to the following (in general complex) ODE for $f(\xi)$ \cite{belanger,akhmediev}:
\eq{Gagnon}{\frac{1}{2}f_{\xi\xi}-(q+b\xi)f+|f|^{2}f-\tau_{R}f(|f|^2)_{\xi}=0,} where $q$ is the wavenumber of the strongest soliton, proportional to its peak power. If one neglects all the nonlinear terms, Eq.(\ref{Gagnon}) would correspond to the stationary Schr\"odinger equation for a unitary mass particle of energy $q$ subject to a 'gravitational' potential $U(\xi) = b\xi$ \cite{gorbach,gorbachnature}. Such gravity-like behavior of the Raman effect in fibers has recently emerged as one of the most important effects in the development of the high-frequency part of the supercontinuum generation (SCG), see Refs. \cite{gorbach,gorbachnature}.

\begin{figure}[htb]
\centerline{\includegraphics[width=8cm]{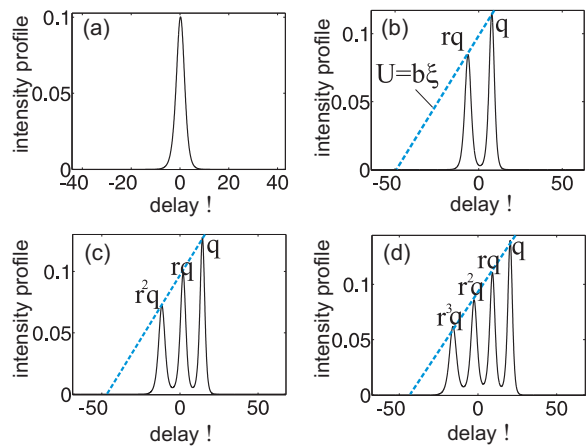}}
\caption{\small  {\bf Profiles of multipeak soliton solutions.} Blue dashed lines show the 'gravity-like' potential $U(\xi)$ created by the most intense soliton in its leading edge into which all other solitons with decreasing powers can be progressively fitted in an 'organ-pipe' fashion.}\label{fig_multi_peaks}
\end{figure}

Numerical solitary-wave solutions of Eq. (\ref{Gagnon}) with one and two peaks were found by Akhmediev for real $f$ \cite{akhmediev}. For any value of the soliton wavenumber $q$, such solutions always show an Airy tail on the leading edge of the pulse, the peak power ratio $r$ and temporal separation $\xi_{0}$ of the maxima in 2-peak solutions being mainly set by material parameters, in particular by $\tau_{R}$. In Ref. \cite{akhmediev}, moreover, it was established that these Raman localised states are metastable, propagating for many tens of dispersion lengths before eventually disappearing due to internal collapse.

In our investigation, we have found a variety of {\em real} solutions of Eq. (\ref{Gagnon}) with an arbitrary number of peaks. In Fig. \ref{fig_multi_peaks} we show the temporal profiles of $1$-, $2$-, $3$- and $4$-peak solitons found by solving the boundary value problem (BVP) by means of a shooting method with appropriate boundary conditions, for $\tau_{R} = 0.1$. Even though these solutions have Airy tails on the leading edge of pulses (due to tunneling of the solutions of the linearized Schr\"odinger equation), such tails are quite small when using the physically relevant parameters, see Fig. \ref{fig_multi_peaks}. Airy tails will be more pronounced if the slope $b$ (or parameter $q$ which is proportional to $\sqrt{b}$) of the gravity-like potential gets larger. The construction of multipeak solutions of Eq. (\ref{Gagnon}) can be understood by looking at the potential term $q+b\xi$ exhibited by Eq. (\ref{Gagnon}). In fact, once the peak power ($2q$) and the position $\xi=\xi_{1}$ of the strongest soliton are given, the second soliton, with peak power $2rq$ ($r \in [0,1]$) and position $\xi=\xi_{2}$, can be arranged only at a temporal separation [see Fig. \ref{fig_multi_peaks}(b)]:
\eq{gravity}{\xi_{0}\equiv\xi_{1}-\xi_{2} \simeq q(1-r)/b.} This simple approximate geometrical reasoning allows one to find a relation between the temporal separation $\xi_{0}$ and the magic peak power ratio $r$.
It is important to note that one can in principle arrange any number of individual solitons into a multipeak state, in an 'organ-pipe' fashion, if the intensity of the strongest soliton is large enough. The strongest soliton creates the gravity-like potential in its leading edge into which all the subsequent solitons, with progressively decreasing peak powers, can be fitted according to the qualitative geometrical relation of Eq. (\ref{gravity}), see Fig.\ref{fig_multi_peaks}.

\begin{figure}[htb]
\centerline{\includegraphics[width=8cm]{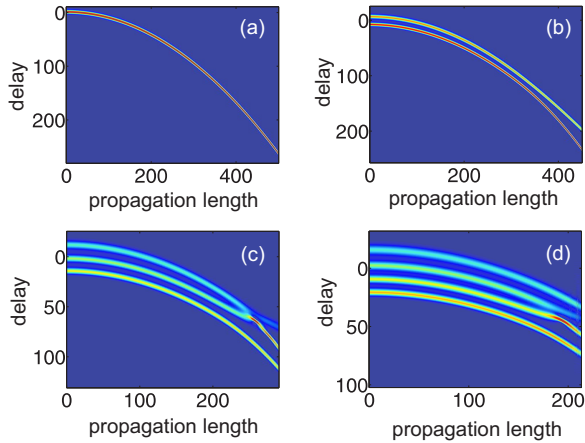}}
\caption{\small  {\bf Temporal evolution of multipeak solitons.}}\label{fig_multipeak_evolution}
\end{figure}

The temporal evolution of 1-, 2-, 3-, and 4- peak solitons is shown in Fig. \ref{fig_multipeak_evolution} (a,b,c,d), respectively, where the corresponding profiles shown in Fig. \ref{fig_multi_peaks} are used as the initial input values for simulation of Eq.(\ref{NLSE}). Apart from the single peak soliton, all other multipeak solutions are not stable, and after some propagation length they will collapse, as clearly shown in Fig. \ref{fig_multipeak_evolution}. The more peaks these localised states have, the less robust they are, and the smaller the propagation length required for them to collapse. In contrast to multipeak solitons, the 1-peak soliton is quite stable, its profile (including its amplitude and width) being stationary to a high degree of approximation, as shown in Fig. \ref{fig_multipeak_evolution}(a).

\subsection{Criteria of 2-peak state formation: magic peak power ratio and temporal separation}
\label{magicratio}

In order to understand the underlying physics that allows the formation of the above Raman localised states we analyze the 2-peak soliton using a Lagrangian method. Two mechanisms are involved in the evolution of 2-peak solitons: (i) the intersoliton forces (ISFs) of two single initially in-phase solitons in the absence of the Raman effect \cite{anderson,karpman}; and (ii) the increasing separation of these two single unequal-amplitude solitons due to RSFS in time domain. Here we treat these two mechanisms independently and {\em require that they have to compensate each other in order to obtain a localised state}. Following a theory formulated in Refs. \cite{anderson,karpman}, due to ISFs two initially in-phase solitons of amplitudes $\sqrt{2q}$ and $\sqrt{2qr}$ with initial temporal separation $\xi_{0}$ in the absence of the Raman effect will pulsate with an oscillation period equal to $L_{\rm p}$:
\eq{Lperiod}{L_{\mathrm{p}}\simeq\frac{\pi}{2(2\nu)^2}\exp\left[\frac{y_0}{4}\right]}
where $\nu = \sqrt{2q}(1+\sqrt{r})/4$, $y_{0} = 4\nu\xi_{0}$. The two solitons have been assumed to have a vanishing phase difference (apart from a constant global phase across the pulse, that can be assumed to be zero due to the gauge invariance of the complex NLSE), since exact numerical soliton solutions of Eq. (\ref{Gagnon}) are always purely real.
\begin{figure}[htb]
\centerline{\includegraphics[width=8cm]{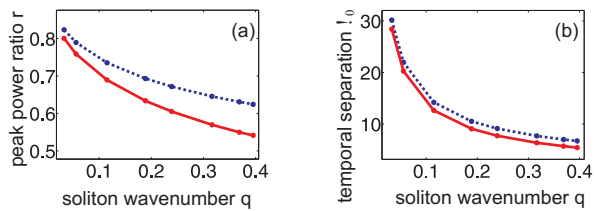}}
\caption{\small  {\bf ({\bf a}) Plot of the peak power ratio $r$ and ({\bf b}) the initial temporal separation $\xi_{0}$ as a function of $q$.} Blue dots and red solid lines indicate the numerical and analytical results, respectively. Parameter $\tau_{R} = 0.1$.}\label{fig_accuracy}
\end{figure}
On the other hand, due to the Raman effect, each single soliton after a propagation length $L_{\rm p}$ will acquire some additional temporal separation $t_{1,2}$ with respect to its own initial position: $t_{1}=\frac{1}{2}bL_{\rm p}^{2} \equiv \frac{16}{15}\tau_{R}q^2L_{\rm p}^{2}$; $t_{2}= \frac{16}{15}\tau_{R}r^2q^2L_{\rm p}^{2}$. Thus, in order to obtain the soliton state where the temporal separation of two single solitons is maintained during propagation, the following condition must hold true: $t_{1}-t_{2} = \xi_{0},$ or:
\eq{analy_curve2}{\pi^2\frac{16}{15}\tau_{R}\frac{1-r^{2}}{(1+\sqrt{r})^{4}}
\exp\left[\frac{1+\sqrt{r}}{\sqrt{2}}\xi_{0}\sqrt{q}\right]=\xi_{0}.}
This condition together with Eq. (\ref{gravity}) leads to a {\em transcendental equation} for the peak power ratio $r$: \eq{transcendental}{\left(\pi\frac{16}{15}\tau_{R}\right)^{2}\frac{2(1+r)q}{(1+\sqrt{r})^{4}}
\exp\left[\frac{15(1+\sqrt{r})(1-r)}{32\tau_{R}\sqrt{2q}}\right]=1.}
From Eqs. (\ref{gravity}) and (\ref{transcendental}) one can easily calculate the peak power ratio $r$ and the temporal separation $\xi_{0}$ for a given wavenumber $q$ of the strongest soliton. Figure \ref{fig_accuracy} shows the dependence of the peak power ratio $r$ and the temporal separation $\xi_{0}$ on the parameter $q$, where red solid lines indicate the analytical (approximate) results, obtained from From Eqs. (\ref{gravity}) and (\ref{transcendental}), whereas the blue dots show the numerical (exact) results, obtained by the shooting method for solving Eq. (\ref{Gagnon}). It is clear from Fig. \ref{fig_accuracy} that the approximate and exact results are in good  qualitative agreement, especially when the wavenumber $q$ of the strongest soliton gets smaller. This behavior is expected, because in this case the slope of the potential $U(\xi)$ shown for the parameter $b \sim q^{2}$ will also get smaller, so that the Airy tail will be negligible. Under these circumstances, the ensemble of two solitons that we use to analytically obtain Eqs. (\ref{Lperiod}) - (\ref{transcendental}) will be closer to the 2-peak numerical solutions with smaller Airy tails.

\subsection{Excitation of 2-peak states}
\label{excitationboundstates}

\begin{figure}[htb]
\centerline{\includegraphics[width=8cm]{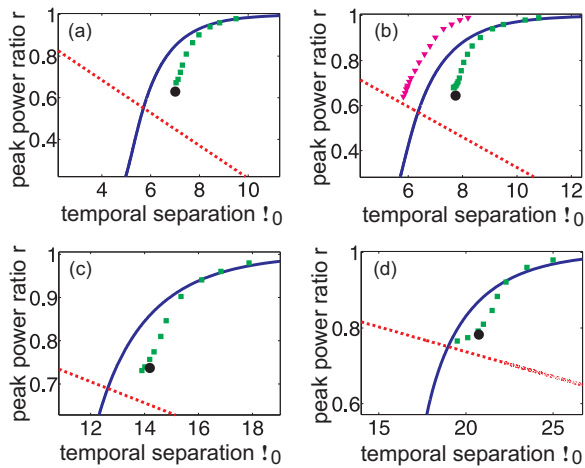}}
\caption{\small {\bf Peak power ratio $r$ versus initial temporal separation $\xi_{0}$.}  ({\bf a,b,c,d}) Results for $q$ = 0.369, 0.3164, 0.1151 and 0.0616, respectively. Red dot lines are calculated based on Eq. (\ref{gravity}), blue solid curves are obtained from Eq. (\ref{analy_curve2}), crossing points of these curves correspond to the solution of Eq. (\ref{transcendental}). Single big black dots are {\em magic points}. Green curves with square markers are obtained by numerically modeling Eq. (\ref{NLSE}) for a propagation length $z$=160 with two $sech$ solitons initially having the same frequency $(\alpha = 0)$ as initial input values. The curve with triangular markers in (b) shows the case $\alpha = -0.2.$}\label{fig_2secant}
\end{figure}

In this Subsection we study the excitation conditions for Raman localised states by numerically integrating Eq. (\ref{NLSE}). As input condition for Eq. (\ref{NLSE}) we use the combination of two hyperbolic secant pulses with an initial frequency detuning  $\alpha$ as follows:
$\Psi_{\rm input}(t)=\sqrt{2q}\mathrm{sech}(\sqrt{2q}t) \exp(-i\alpha t) + \sqrt{2qr} \mathrm{sech}[\sqrt{2qr}(t-\xi_{0})]$. We launch the pulses for a dimensionless propagation length $z = 160$. For each value of the peak power ratio $r$ we need to find the corresponding \emph{initial} temporal separation $\xi_{0}$ such that the temporal separation of the two pulses is as stable as possible during propagation. The results of this numerical analysis are depicted by the green curves with square markers in Fig. \ref{fig_2secant}(a,b,c,d) for four values of $q$ = 0.369, 0.3164, 0.1151 and 0.0616, respectively, for the case when the two solitons have initially the same frequency ($\alpha$ = 0). The big black dots (we refer to these special points ($r$, $\xi_{0}$) as the {\em magic points})  shown in Fig. \ref{fig_2secant} represent the exact numerical soliton solutions of Eq. (\ref{Gagnon}), see also Fig. \ref{fig_multi_peaks}(b).

The red straight lines in Fig. \ref{fig_2secant} are calculated based on Eq. (\ref{gravity}), and the blue solid curves are obtained from Eq. (\ref{analy_curve2}). The crossing points of these curves correspond to the solution of $r$ as found by our qualitative model of Eq. (\ref{transcendental}). The discrepancy between the analytical curves (blue solid ones) and the numerical curves (green curves with square markers) gets smaller for larger $r$ and $\xi_{0}$.

When the parameter $q$ is large enough (see the green curves with square markers in Fig. \ref{fig_2secant}(a,b) with $q$ = 0.369 and 0.3164, for propagation length $z=160$ and initial frequency detuning $\alpha=0$) we do not find any solution ($r$, $\xi_{0}$) very close or below the \emph{magic points} (single big black dots in Fig. \ref{fig_2secant}) so that the temporal separation of the two pulses is preserved during propagation. Of course, the localised states can survive for shorter propagation lengths (e.g. $z \simeq 10$), but they collapse during further propagation. This behavior takes place even exactly at the magic points. This means that the Raman localised states studied here (including magic points) are not mathematically stable, but metastable as found in Ref. \cite{akhmediev}. Nevertheless, these metastable states can propagate for surprisingly long distances, as experimentally observed in Refs. \cite{pod1,pod2}. If we go upwards along the green curves with square markers in Fig. \ref{fig_2secant}, the localised states survive for longer propagation lengths. This is expected, because when the peak power ratio $r$ and the temporal separation $\xi_{0}$ become larger, the amplitude difference between the two pulses falls. As a result, the Raman effect is almost the same for them, they are located further from each other, and thus their ISFs get smaller. In the extreme case when two pulses have the same amplitude and are located infinitely far from each other, this localised state will obviously survive for quite a long propagation length, because each fundamental soliton is very robust \cite{akhmediev}.

The curve with red triangular markers in Fig. \ref{fig_2secant}(b) shows the case when the two pulses have initially different frequencies (the initial frequency detuning $\alpha=-0.2$). In this case, for the same peak power ratio $r$, the \emph{initial} temporal separation $\xi_{0}$ between the two pulses is smaller compared to the case when $\alpha = 0$. But during propagation the temporal separation gets larger and asymptotically stabilizes. In the case $\alpha=0$ the temporal separation is almost the same during the whole propagation length.


\subsection{Magic input powers}
\label{magicinputpowers}

Once the values of $r$ and $\xi_{0}$ have been established, one can explicitly calculate the 'magic input power' of an intense input pulse around which the formation of 2-peak localised states will appear during the pulse splitting process. According to the accepted theory of pulse splitting under the effects of Raman perturbation \cite{splitting1,splitting2}, an intense $N$th-order soliton $\psi_{\rm input}=N\sech{t}$ will split into $N$ individual fundamental solitons of form $\sqrt{P_{j}}\sech{\sqrt{P_{j}}t}$, where the powers $P_{j}$ are given in Eq. (\ref{power}). The condition for which two successive solitons (for instance $j$ and $j+1$) have a peak power ratio equal to $r$, turns into the equation $P_{j+1}/P_{j}=r$, which has the physically relevant solutions $N_{{\rm mag}, j}=(j-1/2)+[1 - \sqrt{r}]^{-1}$. The first two solitons with stronger peak powers ($j=1$ and $j=2$) will have the correct amplitude ratio for $N_{{\rm mag}, 1}=1/2+[1- \sqrt{r}]^{-1}$. For the case $r=R^{2}\simeq0.73$, one obtains $N_{{\rm mag}, 1}\simeq 7.4$ which corresponds surprisingly well to the measured value ($N \simeq 8$), see Refs. \cite{pod1,pod2}.
According to the above straightforward theory, only one 2-peak soliton can be created starting with one intense input pulse, since at each magic input power only one pair of solitons will possess the magic peak power ratio $r$.
It is also clear from the above relation that, for a given value of the peak power ratio $r$,
a higher input soliton order $N$ is required to form a localised state between the $j$-th and the $(j+1)$-th fundamental solitons with higher value of $j$.

\begin{figure}[htb]
\centerline{\includegraphics[width=8cm]{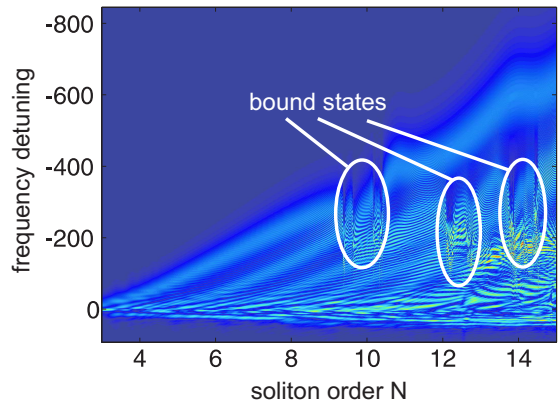}}
\caption{\small {\bf Spectral evolution with increasing soliton number $N$.} The propagation length is $z = 0.45.$}\label{fig_N_v_om}
\end{figure}

Figure \ref{fig_N_v_om} shows the dependence of the spectra at the output for $z$ = 0.45 as a function of the input soliton order $N$ when using a soliton of the form $\psi_{\rm input}=N\sech{t}$ as the initial value for numerical simulation of the generalized NLSE (GNLSE) [Eq. (\ref{GNLSE})] with the full convolution for the nonlinear term:
\eq{GNLSE}{i\de_{z}\psi+\frac{1}{2}\de_{t}^{2}\psi+\psi(z,t) \int_{-\infty}^{t} R(t') |\psi(z,t-t')|^2dt'  =0,} where the response function $R(t)$ includes both the electronic (Kerr) and vibrational (Raman) contributions \cite{agrawalbook}. One has $R(t)= (1-\theta)\delta(t) + \theta h(t)$, where $h(t) \equiv [\tau_{1}^2 + \tau_{2}^2][\tau_{1}\tau_{2}^2]^{-1} \exp(-t/\tau_{2})\sin(t/\tau_{1})$ is the Raman response function of silica, with $\tau_{1}$ = 12.2 fs and $\tau_{2}$ = 32 fs; $\delta(t)$ is the Dirac delta function. Coefficient $\theta$ parameterizes the relative importance between Raman (non-instantaneous) and Kerr (instantaneous) effects, and for silica $\theta \simeq 0.18$. Parameter $\tau_{R}$ in Eqs. (\ref{NLSE}) - (\ref{Gagnon}) is related to $\tau_{1,2}$ as follows: $\tau_{R} = 2\theta \tau_{1}^{2} \tau_{2}/[t_{0}(\tau_{1}^{2} + \tau_{2}^{2})]$, $t_{0}$ being the input pulse duration. With suitable approximations Eq. (\ref{GNLSE}) can be shown to reduce to Eq. (\ref{NLSE}). In Refs. \cite{pod1,pod2} the numerical pulse propagation analysis includes higher order dispersion terms. In this paper, as seen from Eq. (\ref{GNLSE}), we limit ourselves to second order dispersion because we wish to eliminate the phenomenon of resonant radiation emission from solitons \cite{biancalanaresonant}, which was present in the experiments of Refs. \cite{pod1,pod2}, but was not shown there. Operating far from any zero-GVD point increases the chance of observing Raman localised states, since their formation will not be disturbed by the small amplitude background waves generated in the normal dispersion regime close to a zero-GVD point. However, many solid-core PCFs have a zero-GVD in the near-visible and a broad region of anomalous GVD in the far infrared. In such fibers, even though solitons emit radiation if launched near the zero-GVD point, they shift continuously towards the deep anomalous GVD region, where in many cases the GVD is nearly constant for a broad range of frequencies -- hence the constant GVD approximation used in our model of Eq. (\ref{GNLSE}).

\begin{figure}[htb]
\centerline{\includegraphics[width=8cm]{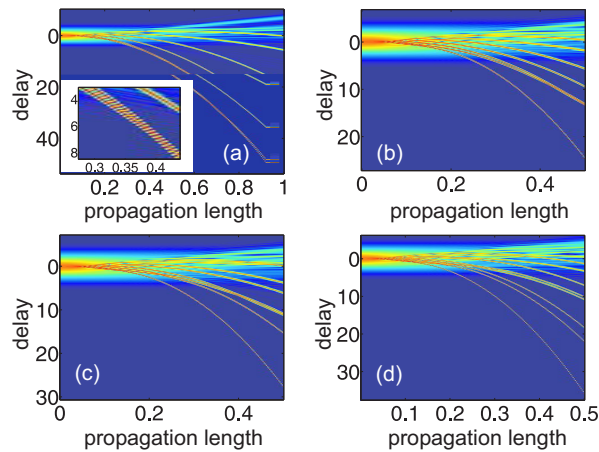}}
\caption{\small {\bf Formation of 2-peak localised states.} ({\bf a,b,c,d}) Temporal evolution of pulses with input soliton order $N$ = 9.63, 12.075, 12.65, and 14.65, respectively. }\label{fig_2peaks}
\end{figure}

Within the three regions enclosed by white ellipses in Fig. \ref{fig_N_v_om} soliton pairs can be formed between adjacent fundamental solitons ejected during the break-up of the input soliton. Four values of soliton order $N$ = 9.63, 12.075, 12.65, and 14.65 are used in these regions to show the pulse temporal evolution in Fig. \ref{fig_2peaks}(a,b,c,d), respectively. The localised pair formed by the 1st and the 2nd fundamental solitons is depicted in Fig. \ref{fig_2peaks}(a) [$N$ = 9.63 is located in the left-hand ellipse in Fig. (\ref{fig_N_v_om})], the inset showing the details of this localised pair and the third separate fundamental soliton. In this case, after a propagation length $z = 1$, the peak power ratio of two pulses forming the Raman state is $r$ = 0.7534, corresponding to the \emph{calculated} value $N_{{\rm mag}}$ = 8.07. Here we should mention that during pulse propagation the two pulses forming the pair will exchange energy, so that the peak power ratio $r$ will also slightly change during propagation. Even though the model used in this section to calculate the magic power is quite crude and qualitative, the correspondence with the experiments is surprisingly good.

Figure \ref{fig_2peaks}(b,c) with input soliton order $N$ = 12.075 and 12.65 (the middle ellipse in Fig. \ref{fig_N_v_om}) shows the states formed by 2nd with 3rd, and 3rd with 4th fundamental solitons, respectively. The pair formed by 4th and 5th fundamental solitons is depicted in Fig. \ref{fig_2peaks}(d) with input soliton order $N$ = 14.65 (the right-hand ellipse in Fig. \ref{fig_N_v_om}). All the localised states shown in Fig. \ref{fig_2peaks} are metastable, collapsing after sufficiently long propagation lengths. The results shown in Fig. \ref{fig_2peaks} are remarkable in the sense that they demonstrate that one can generate pairs for any selected adjacent solitons, simply by properly adjusting the input pulse energy.

In case of solitons represented by magic points, see Figs. \ref{fig_multi_peaks} and \ref{fig_multipeak_evolution}, not only the temporal separation $\xi_{0}$, but also the intensity profiles of each pulse, and thus, the peak power ratio $r$, will be preserved where they are stable. So, the localised states shown in Fig. \ref{fig_2peaks} are not solitons presented by magic points. Moreover, solitons corresponding to magic points are solutions of the simplified model of Eq. (\ref{NLSE}), not of the more complete equation (\ref{GNLSE}).

\begin{figure}[htb]
\centerline{\includegraphics[width=8cm]{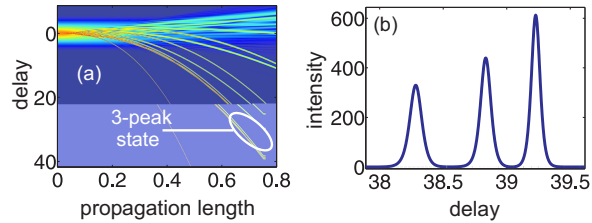}}
\caption{\small {\bf Formation of a 3-peak Raman state.} ({\bf a}) Temporal evolution of a pulse with input soliton order $N$ = 13.81. ({\bf b}) Intensity profile of this 3-peak Raman state at $z$ = 0.8.} \label{fig_3peaks}
\end{figure}

So far we have only investigated the 2-peak states of the GNLSE [Eq. (\ref{GNLSE})]. Apart from 2-peak Raman states, the simplified form of NLSE [Eqs. (\ref{NLSE}-\ref{Gagnon})] also possesses localized solutions with an arbitrary number of peaks [provided that there is enough 'space' to arrange them according to Eq. (\ref{gravity})], as shown in Fig. \ref{fig_multi_peaks}. It turns out that the GNLSE [Eq. (\ref{GNLSE})] also shows evidence of formation of 3-peak states in its dynamics, contrary to the oversimplified conclusions that can be made by using Eqs. (\ref{power}-\ref{width}). This leads to the important prediction that in properly designed realistic PCFs there must be a strong violation of the soliton splitting law given by Eqs. (\ref{power}-\ref{width}). Indeed, Fig. \ref{fig_3peaks}(a) shows the temporal evolution of an intense pulse with the input soliton order $N$ = 13.81 (the right-hand ellipse in Fig. \ref{fig_N_v_om}). During the pulse splitting process, a 3-peak state is formed by the 2nd, 3rd and 4th fundamental solitons. The intensity profile of this 3-peak state at $z$ = 0.8 is shown in Fig. \ref{fig_3peaks}(b). Again, this intensity profile demonstrates that solitons organize themselves in an organ-pipe fashion, as consistent with the gravity-like potential approach used here, see also Fig. \ref{fig_multi_peaks}(c).

\begin{figure}[htb]
\centerline{\includegraphics[width=8cm]{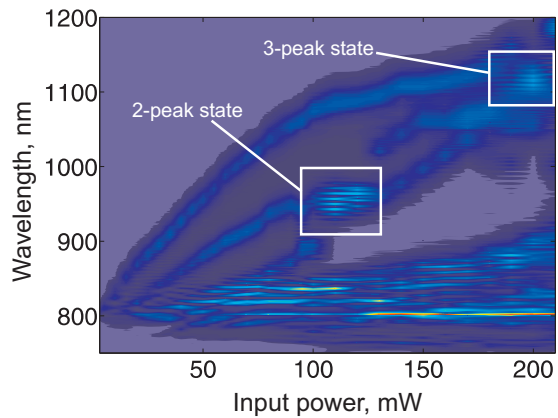}}
\caption{\small  {\bf Experimental data.} Power dependence of the spectra obtained experimentally at the end of a highly nonlinear PCF (18 cm long).} \label{fig_experiment}
\end{figure}

Two-peak Raman states were experimentally observed and reported in Refs. \cite{pod1,pod2}. From the same experimental data we have also observed some indication of the existence of 3-peak states. Figure \ref{fig_experiment} shows the power dependence of the spectra obtained experimentally at the end of a highly nonlinear PCF (18 cm long) \cite{pod1,pod2}. At $\sim$ 100 mW, a 2-peak Raman state is formed by the 2nd and the 3rd fundamental solitons, where at $\sim$ 200 mW one can see some signature of a 3-peak state formed by the 1st, 2nd and 3rd fundamental solitons. These three solitons are close to each other in frequency. At the same time, the presence of strong spectral fringes at $\sim 200$ mW (not shown here) shows that they are also close in time. As mentioned above, 2-peak states are more robust than 3-peak states, but the existence of solitonic states with more than 3 peaks would unambiguously confirm our theoretical model.

\section{Discussion and conclusions}
\label{discussion}

The formation of multipeak localised structures supported by the Raman effect turns out to be a fundamental and surprisingly common process occurring in the region of deep anomalous dispersion in photonic crystal fibers.
The 'gravity-like' potential approach turns out to be a fruitful tool, leading to a simple relationship between two important parameters of 2-peak states, namely the peak power ratio $r$ and the temporal separation $\xi_{0}$ between two pulses forming the pair. Our approach explains in a natural way the 'organ-pipe' arrangement of multipeak states, where the strongest soliton creates a potential in its leading edge and subsequent solitons with monotonously decreasing peak powers can be fitted within this potential, see Fig. \ref{fig_multi_peaks}. The formation of pairs can be explained through the following simple mechanism: the RSFS induces an increasing separation between two unequal-amplitude solitons, that can only be counterbalanced by intersoliton forces. Analyzing these two effects together with the results obtained from the potential approach led us to a simple transcendental equation for the peak power ratio $r$, and thus also for the temporal separation $\xi_{0}$. This transcendental equation gives results that are in good agreement with the exact numerical solutions. The physical interpretation of 2-peak state formation in our model is thus quite simple and intuitive.

We have used a qualitative model to predict the magic pulse power needed to generate a 2-peak state during the pulse splitting process. The predicted values are again in good agreement with experimentally measured ones. In addition to 2-peak states, we have also investigated $n$-peak states with $n>2$. We have demonstrated numerically - using some hints from experiments - that the complexity of the GNLSE allows the existence of 3-peak state. Even though such states are much less robust than 2-peak states, they can survive for at least several tens of dispersion lengths. It is hoped that such 3-peak states can also be experimentally observed in an unambiguous way in the near future.

We expect that the results reported here, apart from their obvious fundamental interest, will allow researchers to manipulate more effectively the dynamics of solitary waves and supercontinuum generation in solid-core microstructured fibers with a properly designed group velocity dispersion, and in hollow-core fibers filled with Raman-active gases.

\section{References}
\label{References}


\section{Acknowledgments}
\label{Acknowledgments}
This work is supported by the German Max Planck Society for the Advancement of Science (MPG).
The authors would like to acknowledge several useful discussions with Alexander Hause and Fedor Mitschke from the University of Rostock, Germany.

\end{document}